# Restricted mean survival time regression model with time-dependent covariates[§]

Chengfeng Zhang[a], Hongji Wu[a], Baoyi Huang[a], Hao Yuan[a], Yawen Hou[b*], Zheng Chen[a*]

[a]Department of Biostatistics, School of Public Health, Southern Medical University,

Guangzhou 510515, China;

[b]Department of Statistics, School of Economics, Jinan University, Guangzhou 510632, China;

* Corresponding author: Zheng Chen and Yawen Hou

Sep. 2021

**Abstract**

In clinical or epidemiological follow-up studies, methods based on time scale indicators such as the restricted mean survival time (RMST) have been developed to some extent. Compared with traditional hazard rate indicator system methods, the RMST is easier to interpret and does not require the proportional hazard assumption. To date, regression models based on the RMST are indirect or direct models of the RMST and baseline covariates. However, time-dependent covariates are becoming increasingly common in follow-up studies. Based on the inverse probability of censoring weighting (IPCW) method, we developed a regression model of the RMST and time-dependent covariates. Through Monte Carlo simulation, we verified the estimation performance of the regression parameters of the proposed model. Compared with the time-dependent Cox model and the fixed (baseline) covariate RMST model, the time-dependent RMST model has a better prediction ability. Finally, an example of heart transplantation was used to verify the above conclusions.

1. Introduction

In clinical follow-up studies, the Cox model is generally used to analyze the relationship between survival outcomes and covariates. The hazard ratio (HR) is a commonly used indicator to measure differences between groups. However, the HR has the following limitations: the HR is a relative measure, whose explanation is not as intuitive as that of an absolute effect. For clinicians, it is difficult to communicate the HR to patients[1]. In addition, when the data do not meet the proportional hazard assumption, the HR will depend on the follow-up duration, and reporting only an HR will lead to incorrect conclusions[2]. Therefore, some researchers have recommended the restricted mean survival time (RMST) as an alternative[3]. Given a specified time point $\tau$, the RMST can be expressed as $\mu(\tau)=E[\min(T,\tau)]$ ($T$ is the survival time), which indicates the average survival time or life expectancy of patients within $[0,\tau]$ [4]. Considering censoring, it can be estimated as the area under the survival curve to a specified time point $\tau$ [5].

Furthermore, considering the relationship between the RMST and covariates, there are currently indirect and direct methods for modeling the RMST and baseline covariates. The main steps of the indirect method are to incorporate the covariates by using different Cox regression models, estimate the corresponding regression coefficients, estimate the cumulative baseline hazard, transform the subject-specific cumulative hazard, and then integrate it to obtain the RMST. Authors, including Kerrison[6], Zucker[7], Chen[8] and Zhang[9], all use different Cox proportional hazard regression models to indirectly estimate the RMST. However, this method is computationally intensive and cumbersome, and it relies on the proportional hazard assumption, which, if untrue, can lead to bias[10]. Hence, several authors have suggested direct methods to model the RMST itself, which mainly include the ANCOVA-type method, pseudo-observation type method, and inverse probability weight (IPW)-type method[4]. Tian's ANCOVA-type method[11] constructed estimating equations for the RMST based on the inverse probability of censoring weighting (IPCW), and the weight function was the inverse of the Kaplan–Meier estimator. Anderson's pseudo-observation method[12-14] modeled the RMST directly based on pseudo-observations. Conner's IPW method[2] adjusted RMST estimators by integrating an adjusted Kaplan-Meier estimator that was adjusted with IPW to obtain propensity scores.

The above regression models all model RMST and baseline covariates; however, in the current clinical follow-up study, time-dependent covariates become increasingly common, and their values will change over time, which is different from baseline covariates. For example, in a heart transplant follow-up trial[15], the endpoint of interest was the death of the patient. As the follow-up time increases, doctors decide whether to perform a heart transplant based on the patient's condition, so the variable of heart transplantation is a time-dependent covariate. The patient's prognosis level is determined by whether heart transplantation is performed. The above regression model cannot handle time-dependent covariates such as heart transplantation.

Therefore, this paper incorporates time-dependent covariates into the model based on IPCW, directly modeling RMST, fixed covariates, and time-dependent covariates. Then, we used two simulations to evaluate the performance of the time-dependent covariate RMST (T-RMST) model: one concerns the performance of the estimator of the regression coefficients in the T-RMST model, and the other concerns the predictive performance of the T-RMST model, which is compared with the time-dependent covariate Cox (T-Cox) model and the fixed-covariate RMST (F-RMST) model. Finally, we provide an illustration by analyzing a heart transplant example.

The structure of the paper is as follows: In Section 2, we describe how to develop a T-RMST model. In Section 3, we describe the two simulations used to evaluate the performance of the T-RMST. In Section 4, we provide an illustration by analyzing a heart transplant example.

## 2. Method

Let $T$ be the survival time for a typical subject; Suppose that $T$ is subject to right censoring by a random variable $C$. Therefore, the observation time is $U = \min(T, C)$, and the event indicator is denoted by $\Delta = I(T \leq C)$, where $I(\cdot)$ is the indicator function. For a time point $\tau$, we denote a fixed covariate $X$ and time-dependent covariate $Z(t)$ $(0 \leq t \leq \tau)$. For subject $i$ ($i$=1, 2, ..., $n$), the observational data consist of $\{U_i, \Delta_i, X_i, Z_i(t)\}$. Let $Y = \min(T, \tau)$ be the restricted survival time, and let $Y_i$ be the corresponding $Y$ for subject $i$. We are interested in the average survival time up to $\tau$ and will model this measure through the fixed covariate and time-dependent covariate:

$$\mu(\tau) = E[Y \mid X, Z(t)].$$

Analogously to a generalized linear model (GLM), we assume a direct relationship between this RMST and the fixed covariate and time-dependent covariate as follows:

$$g(\mu(\tau)) = \boldsymbol{\alpha}^T X^* + \boldsymbol{\beta}^T Z(t). \qquad (1)$$

where $g(\cdot)$ is a link function, $\boldsymbol{\alpha} = (\alpha_0, \alpha_1, ..., \alpha_p)$, $\boldsymbol{\beta} = (\beta_1, \beta_2, ..., \beta_q)$, $X^* = (1, X^T)^T$, and $\alpha_0$ indicates the intercept.

We now derive the estimating equation for the parameters of interest, $\alpha$ and $\beta$. Let $\eta = (\alpha, \beta)$ and $S_i(t) = (X_i^{*T}, Z_i^T(t))^T$. In the absence of censoring, based on (1), $\eta$ can be estimated via the following estimating equation:

$$\frac{1}{n}\sum_{i=1}^{n}\sum_{t \leq \tau} S_i(t)[Y_i - g^{-1}(\eta^T S_i(t))] = 0. \tag{2}$$

However, $E[S_i(t)[Y_i - g^{-1}(\eta^T S_i(t))]] \neq 0$ in the presence of censoring, so the IPCW weighted expectation should still be zero, and $E[S_i(t)W_i^C(t)[Y_i - g^{-1}(\eta^T S_i(t))]] = 0$. Then, the estimated equation changes from (2) to:

$$\frac{1}{n}\sum_{i=1}^{n}\sum_{t \leq \tau} S_i(t)\tilde{\Delta}_i W_i^C(t)[Y_i - g^{-1}(\eta^T S_i(t))] = 0, \tag{3}$$

where $\tilde{\Delta}_i = I(Y_i \leq C_i)$, $W_i^C(t) = W_i^X(t)W_i^Z(t)$, $W_i^X(t) = \exp(H_i^X(t))$, and $W_i^Z(t) = \exp(H_i^Z(t))$. $H_i^X(t)$ and $H_i^Z(t)$ are the cumulative hazards of fixed covariates and time-dependent covariates under the censored time distribution. In general, $H_i^X(t)$ and $H_i^Z(t)$ are rarely known in practice and therefore must be estimated from the observed data. Therefore, based on the censoring time distribution, we use the Cox model and the T-Cox model to calculate $H_i^X(t)$ and $H_i^Z(t)$, respectively, and we fit the corresponding models:

$$h_i^X = h_0^X(t)\exp(\alpha_C^T X_i),$$

$$h_i^Z = h_0^Z(t)\exp(\beta_C^T Z_i(t)).$$

Thus, we calculate the cumulative hazard by $\hat{H}_i^X(t) = \int_0^t \hat{h}_i^X(u)du$ and $\hat{H}_i^Z(t) = \int_0^t \hat{h}_i^Z(u)du$. Plugging $\hat{H}_i^X(t)$ and $\hat{H}_i^Z(t)$ into (3), we can obtain the following estimating equation:

$$\frac{1}{n}\sum_{i=1}^n \sum_{t\leq\tau} S_i(t)\tilde{\Delta}_i \hat{W}_i^C(t)[Y_i - g^{-1}(\eta^T S_i(t))] = 0, \quad (4)$$

where $\hat{W}_i^C(t) = \hat{W}_i^X(t)\hat{W}_i^Z(t)$, $\hat{W}_i^X(t) = \exp(\hat{H}_i^X(t))$, and $\hat{W}_i^Z(t) = \exp(\hat{H}_i^Z(t))$. $\eta$ can be estimated from estimation equation (4).

According to Tian[11] and Zhong[10], we have $\sqrt{n}(\hat{\eta}-\eta) \sim N(0, A^{-1}BA^{-1})$; therefore, $\hat{V}(\hat{\eta}) = \hat{A}^{-1}\hat{B}\hat{A}^{-1}$, where

$$\hat{A} = E[\sum_{t\leq\tau} S_i(t)^{\otimes 2} \dot{g}^{-1}(\hat{\eta}^T S_i(t))],$$

$$\hat{B} = E[\sum_{t\leq\tau} \varepsilon_i(\hat{\eta})^{\otimes 2}],$$

$$\varepsilon_i(\hat{\eta}) = S_i(t)\tilde{\Delta}_i \hat{W}_i(t)[Y_i - g^{-1}(\hat{\eta}^T S_i(t))].$$

$a^{\otimes 2} = aa^T$ for vector $a$, and $\dot{g}^{-1}(\cdot)$ is the derivative of $g^{-1}(\cdot)$. The asymptotic standard error (ASE) can be estimated from $ASE = \sqrt{\frac{1}{n}diag(\hat{V}(\hat{\eta}))}$.

Using the above model (1), one may estimate $\mu(\tau)$ by $\hat{\mu}(\tau) = g^{-1}(\hat{\alpha}^T \tilde{X}^* + \hat{\beta}^T \tilde{Z}(t))$, where $X = \tilde{X}$, $\tilde{X}^* = c(1, \tilde{X})$, and $Z_i(t) = \tilde{Z}_i(t)$. Let $\hat{\eta} = (\hat{\alpha}, \hat{\beta})$ and $\tilde{S}(t) = (\tilde{X}^{*T}, Z^T(t))^T$; thus, the 95% confidence interval of $\hat{\mu}(\tau)$ is $\hat{\mu}(\tau) \pm t_{0.025,v} SE$, where $v$ is the degree of freedom of the regression equation and *SE* can be estimated by the delta method.

$$SE = \sqrt{\widehat{Var}(g^{-1}(\hat{\alpha}^T \tilde{X}^* + \hat{\beta}^T \tilde{Z}(t)))}$$
$$= \sqrt{\widehat{Var}(g^{-1}(\hat{\eta}^T \tilde{S}(t))}$$
$$= \sqrt{\left(\frac{\partial(g^{-1}(\hat{\eta}^T \tilde{S}(t)))}{\partial\hat{\eta}}\right)^T \widehat{Var}(\hat{\eta})\left(\frac{\partial(g^{-1}(\hat{\eta}^T \tilde{S}(t)))}{\partial\hat{\eta}}\right)},$$

where

$$\frac{\partial(g^{-1}(\hat{\boldsymbol{\eta}}^T\tilde{\boldsymbol{S}}(t)))}{\partial\hat{\boldsymbol{\eta}}} = \left(\frac{\partial(g^{-1}(x))}{\partial x}\right)^2_{x=\hat{\boldsymbol{\eta}}^T\tilde{\boldsymbol{S}}(t)} \frac{\partial(\hat{\boldsymbol{\eta}}^T\tilde{\boldsymbol{S}}(t))}{\partial\hat{\boldsymbol{\eta}}}$$

$$= \left(\frac{\partial(g^{-1}(x))}{\partial x}\right)^2_{x=\hat{\boldsymbol{\eta}}^T\tilde{\boldsymbol{S}}(t)} \tilde{\boldsymbol{S}}(t)$$

## 3. Simulation

### 3.1 Regression coefficients

#### 3.1.1 Simulation design

We used Monte Carlo simulation to evaluate the regression coefficient estimation effect of the proposed T-RMST model. The time-dependent covariate can be represented as $Z(t)=0$ for $t<t_0$, while $Z(t)=1$ for $t\geq t_0$, where $t_0 \sim U(0,4)$. Suppose that event times follow a Weibull distribution; therefore, we simulated the survival times as

$$T = \begin{cases} \left(\frac{-\log(u)}{\lambda\exp(\beta'X)}\right)^{1/\nu}, & -\log(u) < \lambda\exp(\beta'X)t_0^\nu \\ \left(\frac{-\log(u)-\lambda\exp(\beta'X)t_0^\nu+\lambda\exp(\beta_t)\exp(\beta'X)t_0^\nu}{\lambda\exp(\beta_t)\exp(\beta'X)}\right)^{1/\nu}, & -\log(u) \geq \lambda\exp(\beta'X)t_0^\nu \end{cases},$$

where $u \sim U(0,1)$ [16]. Fixed covariates were sampled from a Bernoulli (0.5) distribution. The time-dependent covariate regression coefficient ($\beta_t$) and fixed covariate regression coefficient ($\beta'$) were both set to 0.1. We also set the shape ($\lambda$) and scale ($\nu$) parameters to 0.1 and 1.5, respectively. The censoring times were generated from an exponential distribution, resulting in different censoring rates. The detailed steps are as follows:

1) To mimic the population, we generated a dataset with a number of subjects $N=1000000$ under approximately zero censoring[17, 18]. The T-RMST model was fitted by the dataset, and the "true" regression coefficients ($\beta_0, \beta_1, \beta_2$) could be calculated by the equation: $g(\mu(\tau)) = \beta_0 + \beta_1 X + \beta_2 Z(t)$.

2) We generated datasets with a number of subjects $N=1000000$ under approximately 15, 30 and 45% censoring, sampled from each dataset, fit the T-RMST model, and calculated the corresponding regression coefficients and variances of the samples. We repeated the above steps 10000 times.

3) We compared the regression coefficients obtained from the sample with the "true" regression coefficients ($\beta_0, \beta_1, \beta_2$), and we used the bias, mean square error (MSE), relative standard error (Rel SE) and empirical coverage probabilities (CP) to evaluate the regression coefficient estimation effect [19, 20].

### 3.1.2 Simulation results

Table 1 shows the estimation effect of the regression coefficients for sample sizes $n=500$ and $n=1000$ under different censoring rates. It can be seen that the bias of the regression coefficients calculated by the sample was small; most of the bias values were less than 1%, and the maximum was not more than 7%. The MSE was also relatively small; most of the MSE values were below 20%, the maximum was approximately 37%, and they decreased with increasing sample size and decreasing censoring. Additionally, the Rel SE was approximately equal to 1, and the CP was similarly very close to the nominal level, indicating that the regression coefficients were well estimated.

### 3.2 Predictive performance

**3.2.1 Simulation design**

Similarly, Monte Carlo simulation was used to evaluate the predictive performance of the proposed T-RMST model. We generated sample sizes of 500 and 1000, where the training set was 2/3 of the sample and the test set contained the remaining samples. The same method of simulating the survival time was used as in Section 3.1.1. The parameter settings were also the same as in Section 3.1.1. The censoring times were generated from an independent exponential distribution, resulting in approximately 15, 30, and 45% censoring rates. The T-Cox, T-RMST and F-RMST [11] models were compared under different sample sizes and censoring rates. All simulations were performed using 1000 iterations. The predictive performances of the different models were evaluated by Harrell's C-index[21] and the prediction error[11, 22]. The C-index measures the probability of concordance between the predicted order and the observed order, and the prediction error is the difference between the predicted value and the true value. A higher C-index and lower prediction error indicate a better-performing model, and the prediction effect of the model is evaluated by these two indicators.

**3.2.2  Simulation results**

Table 2 shows the prediction effects of the T-Cox, T-RMST, and F-RMST models under different sample sizes and different censoring rates. The results obtained with different sample sizes and censoring rates are basically consistent. The C-index of the T-RMST model is higher than those of the T-Cox and F-RMST models. The prediction error of the T-RMST model is lower than that of the F-RMST model.

Combining these two indicators, it can be seen that the T-RMST model enjoys a better prediction performance.

**4. Example**

The data comes from the Stanford Heart Transplant Center and includes 103 patients. The outcome of this analysis was overall survival, which was calculated in years from the time of diagnosis to death[15]. Patients who were still alive at the last follow-up were censored. Intermediate events such as heart transplant occurred during the follow-up period, the longest of which was 4.93 years; in other words, during the follow-up period, the doctor decided whether to perform a heart transplant based on the patient's condition. The fixed covariates included age (which was divided into young people (<45), middle-aged people (45–60), and elderly people ($\geq 60$) according to the World Health Organization (WHO) division rules), the patient's enrollment time (the time of enrolling in the project minus the study start time 1967/10/01) and whether the patient had had heart bypass surgery. Since one patient in the data died on the day of entry, to enable the T-Cox model to handle such data, the survival time of this patient was increased to $0.5$[23]. Then, we applied the T-Cox, T-RMST and F-RMST models to obtain the regression results.

The analysis results of the three models are shown in Table 3. The differences in RMST (RMSTd) of patients who were younger than 45 and older than 60 were -0.766 years (95% CI: -1.345, -0.187) and -1.047 years (95% CI: -1.692, -0.402) with the T-RMST and F-RMST models, respectively. However, the results of the T-Cox model showed that it was not statistically significant for the prognosis of patients. When the enrollment time was longer than 1 year, the RMSTd values were -0.149 years (95% CI: -0.290, -0.008) and -0.237 years (95% CI: -0.378, -0.095) with the T-RMST and F-RMST models, respectively. However, the results of the T-Cox model showed that the HR was 0.855 (95% CI: 0.745, 0.980). The patients who underwent heart bypass surgery survived 0.778 years (95% CI: 0.115, 1.441) and 1.189 years (95% CI: 0.409, 1.969) longer than those who did not according to the T-RMST and F-RMST models, respectively. However, the results of the T-Cox model showed that it was not statistically significant for the prognosis of patients. In addition, for the time-dependent covariate of heart transplantation, the T-RMST model showed that the RMSTd was 0.868 years (95% CI: 0.407,1.328) and that the patients could live 0.868 years longer after transplantation, while the T-Cox model showed that heart transplantation was not statistically significant. At the same time, the $R^2$ values of T-Cox (discrete model) [24], T-RMST and F-RMST were calculated. The larger the value of $R^2$ is, the better the fitting effect of the model, and the values were 0.075, 0.329 and 0.187.

We used the C-index and prediction error to evaluate the predictive performance of the model. First, the instance dataset was divided into a training set and a test set, where the training set was a random sample of 2/3 of the data and the test set contained the remaining samples. Then, three models that selected meaningful covariates were fitted to the training set. Finally, the C-indexes of the three models were calculated as well as the prediction error of the T-RMST and F-RMST models; the above steps were repeated 500 times to obtain the average C-index and prediction error. The results are shown in Table 4. The C-indexes of T-Cox, T-RMST and F-RMST were 0.541, 0.652, and 0.531, respectively, and the prediction errors of T-RMST and F-RMST were 0.337 and 1.085, respectively.

Finally, the T-RMST model can predict the average survival time of the patient in the next 4.93 years according to the individual characteristics of the patient [25]. For example, for a patient aged 40 years enrolled one year previously who had received heart bypass surgery but not a heart transplant, the average survival time for the next 4.93 years was 1.055 years (95% CI: 0.238, 1.872). When the patient received a heart transplant, the average survival time for the next 4.93 years was 1.923 years (95% CI: 0.985, 2.861).

## 5. Discussion

Time-dependent covariates are becoming increasingly common in clinical follow-up studies. Therefore, this paper developed the T-RMST regression model based on IPCW processing for time-dependent covariates. The IPCW weight is not the inverse of the survival rate estimated by Kaplan-Meier analysis but the double inverse weight required and estimated through the Cox model and the T-Cox model for fixed covariates and time-dependent covariates. Because there are time-dependent covariates, the survival curve changes during follow-up, so it is biased to directly use the standard Kaplan-Meier method to calculate the survival rate and then convert it into weights, which will overestimate the variance of the Kaplan-Meier distribution. The simulation results of the regression coefficients showed that under different sample sizes and censorship rates, combined with the four indicators of bias, MSE,

Rel SE and CP, taking into account time-dependent covariates, the regression coefficient estimation performance of the T-RMST model was better. The predictive performance simulation results showed that considering the C-index and prediction error of the two predictive indicators together, the T-RMST had better predictive performance than the T-Cox and F-RMST models.

In the example of heart transplantation, combining the three predictive indicators of $R^2$, C-index and prediction error, the T-RMST model had the best predictive effect. Moreover, the T-RMST model and the F-RMST model had the same results in dealing with fixed covariates, which were different from those of the T-Cox model. According to a paper by Khush[26] on heart transplantation, an increase in age will reduce the survival rate of heart transplant patients, and receiving a heart transplant will increase the survival rate of patients and prolong the survival time. This is consistent with the conclusions obtained by the T-RMST and the F-RMST models; thus, the reliability of the conclusions of the T-RMST model is confirmed from a clinical perspective. We can also use the T-RMST model to predict the RMST value based on individual characteristics, especially the existence of time-dependent covariates.

The T-RMST model also has some limitations. First, the IPCW weights are unstable. The existence of extreme values may lead to weights that are too large or too small, leading to extreme values of the regression coefficients and their variances, resulting in the stability of the predicted RMST values; however, the weights can be stabilized by setting an appropriate $\tau$ [10]. In addition, the T-RMST model is not capable of processing the endogenous time-dependent covariates, but the dynamic

RMST model can do this[27].

In summary, the T-RMST model developed in this paper is a regression model that takes into account time-dependent covariates, and its predictive effect is better than that of the traditional T-Cox model. In addition, because the RMST is not required to satisfy the assumption of proportional hazards in the data, the T-RMST model can handle a wider range of data types. At the same time, RMSTd explains the nature of covariates on a time scale. For example, for a patient who receives a heart transplant, it is easier for him or her to understand "how long I can live" than "how much I have reduced the risk of death", so this measure is easier to understand than HR. Compared with the traditional fixed-covariate F-RMST model, the T-RMST model can better deal with time-dependent covariates, and the prediction effect is also better than that of the former.

**Funding**: This work was supported by the National Natural Science Foundation of China [grant numbers 82173622, 81903411, 81673268] and the Guangdong Basic and Applied Basic Research Foundation [grant number 2019A1515011506].

**Conflict of Interest**: The authors declare no conflict of interest.

**Data Sharing:** The data that support the findings of this study are openly available in R package survival at https://CRAN.R-project.org/package=survival, reference number 15.

**Table 1** Performances of the estimators of the regression coefficients in T-RMST

| n | Cen (%) | Coef | True | Bias | MSE | Rel SE | CP |
|---|---|---|---|---|---|---|---|
| 500 | 15 | $\beta_0$ | 1.6439 | -0.0015 | 0.1441 | 1.0100 | 0.9509 |
|  |  | $\beta_1$ | -0.1885 | 0.0038 | 0.2264 | 1.0185 | 0.9534 |
|  |  | $\beta_2$ | 3.1630 | 0.0136 | 0.1684 | 1.0326 | 0.9552 |
|  | 30 | $\beta_0$ | 1.6439 | -0.0014 | 0.1690 | 1.0178 | 0.9532 |
|  |  | $\beta_1$ | -0.1885 | 0.0079 | 0.2798 | 1.0183 | 0.9502 |
|  |  | $\beta_2$ | 3.1630 | 0.0042 | 0.1909 | 1.0889 | 0.9618 |
|  | 45 | $\beta_0$ | 1.6439 | -0.0005 | 0.2135 | 1.0029 | 0.9510 |
|  |  | $\beta_1$ | -0.1885 | 0.0271 | 0.3692 | 0.9991 | 0.9410 |
|  |  | $\beta_2$ | 3.1630 | -0.0661 | 0.2370 | 1.1560 | 0.9454 |
| 1000 | 15 | $\beta_0$ | 1.6439 | -0.0006 | 0.1000 | 1.0283 | 0.9583 |
|  |  | $\beta_1$ | -0.1885 | 0.0015 | 0.1598 | 1.0225 | 0.9541 |
|  |  | $\beta_2$ | 3.1630 | 0.0152 | 0.1185 | 1.0433 | 0.9589 |
|  | 30 | $\beta_0$ | 1.6439 | -0.0007 | 0.1202 | 1.0199 | 0.9537 |
|  |  | $\beta_1$ | -0.1885 | 0.0066 | 0.2002 | 1.0213 | 0.9533 |
|  |  | $\beta_2$ | 3.1630 | 0.0134 | 0.1336 | 1.1171 | 0.9677 |
|  | 45 | $\beta_0$ | 1.6439 | 0.0031 | 0.1540 | 1.0195 | 0.9527 |
|  |  | $\beta_1$ | -0.1885 | 0.0198 | 0.2711 | 1.0113 | 0.9492 |
|  |  | $\beta_2$ | 3.1630 | -0.0376 | 0.1615 | 1.2346 | 0.9674 |

Note: *n*, the sample size; Cen, the censoring rate; Abbreviations: Coef, Coefficient.

Note: Bias, $E(\hat{\beta}) - \beta$; MSE, mean square error, $E(\hat{\beta} - \beta)^2$; Rel SE, asymptotic standard error/empirical standard deviation $ASE/ESD$; CP, coverage probabilities, $\Pr(\hat{\beta}_{low} \leq \beta \leq \hat{\beta}_{upp})$.

**Table 2** Prediction performances of T-Cox, T-RMST and F-RMST

| n | Cen (%) | T-Cox | | T-RMST | | F-RMST | |
|---|---|---|---|---|---|---|---|
| | | C-index | PE | C-index | PE | C-index | PE |
| 500 | 15 | 0.440 | \ | 0.669 | 0.985 | 0.506 | 2.036 |
| | 30 | 0.441 | \ | 0.675 | 1.031 | 0.504 | 2.029 |
| | 45 | 0.446 | \ | 0.679 | 1.066 | 0.505 | 2.012 |
| 1000 | 15 | 0.427 | \ | 0.672 | 0.983 | 0.510 | 2.025 |
| | 30 | 0.425 | \ | 0.676 | 1.037 | 0.508 | 2.027 |
| | 45 | 0.424 | \ | 0.680 | 1.087 | 0.507 | 2.015 |

Note: *n*, the sample size; Cen, the censoring rate; Abbreviations: PE: Prediction error

Note: A higher C-index indicates a better-performing model; a lower prediction error indicates a better-performing model.

**Table 3** The results of the T-Cox, T-RMST and F-RMST models in the example

| Variable | T-Cox | | | T-RMST | | | F-RMST | | |
|---|---|---|---|---|---|---|---|---|---|
| | HR | 95%CI | *P* | RMSTd | 95%CI | *P* | RMSTd | 95%CI | *P* |
| **Intercept** | | | | 0.426 | (-0.031,0.882) | 0.068 | 1.770 | (0.888,2.651) | <0.001 |
| **Age** | Ref: age<45 (years) | | | | | | | | |
| 45–60 | 1.397 | (0.851,2.290) | 0.186 | 0.042 | (-0.459, 0.544) | 0.869 | -0.277 | (-0.985,0.432) | 0.444 |
| >=60 | 1.881 | (0.432,8.200) | 0.400 | -0.766 | (-1.345,-0.187) | 0.010 | -1.047 | (-1.692,-0.402) | 0.001 |
| **Enrollment time** | 0.855 | (0.745,0.980) | 0.025 | -0.149 | (-0.290,-0.008) | 0.038 | -0.237 | (-0.378,-0.095) | 0.001 |
| **Bypass surgery** | Ref: No | | | | | | | | |
| Yes | 0.515 | (0.249,1.060) | 0.073 | 0.778 | (0.115,1.441) | 0.021 | 1.189 | (0.409,1.969) | 0.003 |
| **Transplant** | 1.077 | (0.590,1.960) | 0.809 | 0.868 | (0.407,1.328) | <0.001 | | | |

Abbreviations: RMSTd: The difference in the restricted mean survival time. Coef: Coefficient.

Note: In the T-Cox model, HR=exp(Coef); In the T-RMST or F-RMST model, RMSTd=Coef.

**Table 4** C-indexes and prediction errors of the T-Cox, T-RMST and F-RMST models in the example

|  | **T-Cox** | **T-RMST** | **F-RMST** |
|---|---|---|---|
| C-index | 0.541 | 0.652 | 0.531 |
| Prediction error | \ | 0.337 | 1.085 |